\documentclass[twocolumn,prb,twoside,a4paper]{revtex4}
\pdfoutput=1

\usepackage[font=footnotesize,labelfont=sf,bf,flushleft]{caption}
\usepackage{graphicx}
\usepackage[usenames,dvipsnames]{xcolor}
\usepackage{amsmath,bbm}
\usepackage{natbib}
\usepackage[normalem]{ulem}
\usepackage{titlesec}
\usepackage{multirow}
\titlespacing{\section}{0pt}{0pt}{-4pt}
\titlespacing{\subsection}{0pt}{0pt}{-4pt}

\titleformat{name=\section}{\filright \sf \normalsize \bfseries}{\thesection}{.5em}{}
\titleformat{name=\subsection}{\filright \sf \small \bfseries}{\thesubsection}{.5em}{}

\newcommand{\shorteq}{%
  \settowidth{\@tempdima}{-}
  \resizebox{\@tempdima}{\height}{=}%
}

\begin{document}

\title{\flushleft{ \Huge{\bfseries A levitated nanocryostat: Laser refrigeration, alignment and rotation of Yb$^{3+}$:YLF
nanocrystals}}}

\author{\hspace{-2.2cm} \bf A. T. M. Anishur Rahman \& P. F. Barker$^*$} 
\affiliation{\footnotetext{Department of Physics $\&$ Astronomy, University College London, Gower Street, WC1E 6BT, London, UK. $^*$E-mail: p.barker@ucl.ac.uk.}}


\maketitle

\renewcommand{\baselinestretch}{1}

\small 
\noindent {\bfseries}


\textbf{The ability to cool and manipulate levitated nano-particles in vacuum is a promising new tool for exploring macroscopic quantum mechanics\cite{WanPRL2016,Scala2013,Zhang2013}, precision measurements of forces, \cite{GambhirPRA2016} and non-equilibrium thermodynamics \cite{GieselerNatNano2014,MillenNat2014}. The extreme isolation afforded by optical levitation offers a low noise, undamped environment that has to date been used to measure zeptonewton forces \cite{GambhirPRA2016}, radiation pressure shot noise,\cite{Jain2016} and to demonstrate the cooling of the centre-of-mass motion \cite{LiNatPhys2011,Gieseler2012}.  Ground state cooling, and the creation and measurement of macroscopic quantum superpositions, are now within reach, but control of both the center-of-mass and internal temperature is required. While cooling the centre-of-mass motion to microKelvin temperatures has now been achieved, the internal temperature has remained at or well above room temperature. Here we demonstrate refrigeration of levitated Yb$^{3+}$:YLF from room temperature to 130 K using anti-Stokes fluorescence cooling, while simultaneously using the optical trapping field to align the crystal to maximise cooling. }

\begin{figure*}
\includegraphics{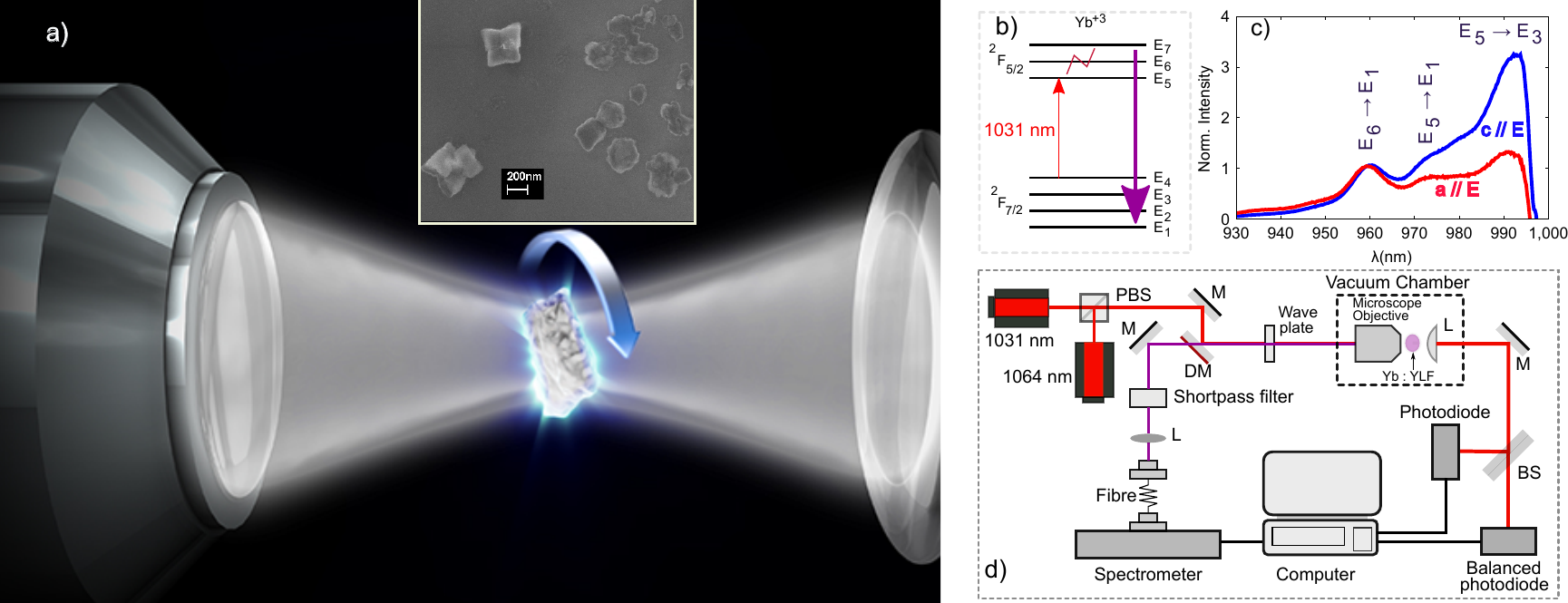}
\caption{
\footnotesize{\textbf{Experimental overview.}}
\textbf{ a)} A illustration of the levitated Yb$^{3+}$:YLF crystal that is excited by the trapping beam and emitted blue-shifted fluorescence. Inset is a scanning electron microscope image of the nanoparticles used in the experiment. The scale bar is $200$ nm. \textbf{ b)} The energy level diagram of the Yb$^{3+}$ ion within the YLF crystal. Shown is the excitation at $1031$ nm and the anti-stokes fluorescence at shorter wavelengths which lead to internal cooling of the crystal. \textbf{ c)} Fluorescence spectra obtained from a levitated Yb$^{3+}$:YLF crystal under $1031$ nm illumination- blue and red solid lines represent electric field $E$ of the trapping laser parallel to the crystal $c$-axis and $a-$ axis, respectively. \textbf{d) } Optical layout of the experiment - M denotes a mirror, DM stands for a dichroic mirror, PBS means a polarizing beam splitter, BS denotes a beam splitter, and L signifies a lens.}
\label{fig1} 
\end{figure*}

Laser cooling the centre-of-mass motion of atoms to ultra-cold temperatures has led to a revolution in physics over the last thirty years. This has allowed the creation of quantum gases, the study of quantum many body physics and novel states of matter. It has become a workhorse in atomic and quantum physics and is now being developed as an important tool in quantum technology and metrology. In more recent years cooling of more complex systems including molecules and even levitated nanoparticles has been demonstrated.
 
Laser cooling (refrigeration) of solid materials has, despite first predictions \cite{Pringsheim1929} in 1929, made somewhat slower progress. Like the well known process of Doppler cooling \cite{Metcalf2007}, blue shifted spontaneous emission of photons with respect to the excitation wavelength leads to cooling. The difference is that unlike Doppler cooling of atoms, it is the internal temperature of the solid that is lowered by converting phonons to higher energy photons, thereby decreasing both the temperature and entropy of the system\cite{Mungan1997,Seletskiy2010}. Last year cryogenic temperatures ($90$ K) were realised for the first time using this technique \cite{Melgaard2016} in a bulk crystal of Yb$^{3+}$:YLF, while the first demonstration \cite{Epstein1995} in $1995$ cooled Yb:ZBLAN glass by $0.3$ K. The lowest temperature that can be reached is believed to be limited by parasitic trace impurities in the solid which absorb the cooling light and lead to heating. 

This type of cooling opens up a wide range of applications, where in contrast to conventional cryogenic methods, heat transfer from the solid to the environment occurs by fluorescence emission \textit{without} physical contact. Such a mechanism now offers cryogenic cooling within enclosed spaces without the need for electrical connections or conductive materials, with clear applications within a vacuum or space environment\cite{Seletskiy2010}.  Recently, laser refrigeration on a microscopic scale has been demonstrated in optical tweezers experiments using Yb$^{3+}$:YLF crystals in water\cite{Roder2015}. Here the water temperature was reduced by $11$ K,  with the final temperature limited by the high convective heat transfer of liquid water. Refrigeration on the microscale in such biophysical environments is seen as a promising tool for exploring the temperature effects of physiological processes.

In this letter we report on the refrigeration of levitated Yb$^{3+}$:YLF nanocrystals to temperatures as low as 130 K using a single beam optical dipole trap (optical tweezers) within a low pressure environment. The same light field that traps also leads to anti-stokes refrigeration of the crystal. The temperature of the crystal is observed by both measuring the changes in fluorescence emitted from the nanocrystal as a function of laser wavelength, and also by the temperature-dependent damping of the motion of the nanocrystal in the trap. We show that temperature can be controlled over a wide range (130 - 300 K) using different trapping laser wavelengths, while the laser polarization allows us to control the orientation of the trapped crystal and maximise its cooling.

A schematic of our experiment for levitated anti-stokes refrigeration is shown in Fig. \ref{fig1}a. A laser beam is tightly focussed into a diffraction limited spot by a $0.80$ numerical aperture microscope objective to form a $3$D optical dipole trap. The objective lens is held inside a vacuum chamber in which the gas can be evacuated once the particle is trapped at atmospheric pressure. This same laser acts to both levitate and refrigerate the Yb$^{3+}$:YLF nanoparticles. The energy level diagram of Yb$^{3+}$ in YLF is shown in Fig. \ref{fig1}b, while panel c shows the recorded blue-shifted fluorescence, which leads to cooling, from a levitated Yb$^{3+}$:YLF nanoparticle. The fluorescence is collected using the same trapping objective and is sent to a spectrometer. A lens (Fig. \ref{fig1}d) collects both the transmitted and scattered light from a levitated nanocrystal, which is then directed to photodiodes that measure the rotational and translational motion of the trapped crystal.

The energy level structure of Yb$^{3+}$ embedded in the yttrium lithium fluoride consists of an $F_{7/2}$ ground state and a $F_{5/2}$ excited state, each of which is split into a manifold of non-degenerate states by the crystal field \cite{Bensalah2004,Sugiyama2006}. The radiative lifetime of the excited state in the bulk crystal is on the order of $2$ milliseconds\cite{Bensalah2004,Fan2007} and there is no significant non-radiative relaxation \cite{Seletskiy2010,Seletskiy2013}. In contrast, rapid phonon coupling between the crystal-split manifold levels in each electronic state occurs over picosecond time-scales, rapidly thermalising the population in these levels.  Anti-stokes refrigeration occurs via optical excitation at a wavelength that is resonant with the bandgap between the two electronic states at approximately 1020 nm (as shown in figure 1b). Due to the rapid thermalisation in the upper state and the long excited state lifetime, the fluorescence from the excited state is blue-shifted with respect to the excitation wavelength. This reduces the average phonon energy and the lattice temperature of the levitated Yb$^{3+}$:YLF nanocrystal.

Figure \ref{fig1}c shows the spectrally dispersed fluorescence from a levitated Yb$^{3+}$:YLF nanocrystal using linearly polarised trapping light over the spectral region from 930-995 nm for a trap formed by 1031 nm light. Two types of spectral profiles are observed. When trapping at high laser powers ($100-300$ mW), the typical spectral profile observed is shown by the blue curve. This profile corresponds to the crystal's $c$-axis oriented parallel to the polarization of the optical field(blue lines in figure).  In this orientation the maximum absorption and cooling has been observed in bulk crystals. When the trapping power is lowered to below $80$ mW, the spectral profile changes to that shown by red curve. This profile corresponds to an orientation of the crystal's $a$-axis parallel to the laser polarisation. This well defined alignment as a function of laser power is in contrast to the random alignment of the crystal axes when the nanocrystals are deposited on a microscope cover slip. This type of alignment has been observed in the trapping of birefringent crystals in aqueous media\cite{Singer2006}.

\begin{figure*}
\centering
\includegraphics{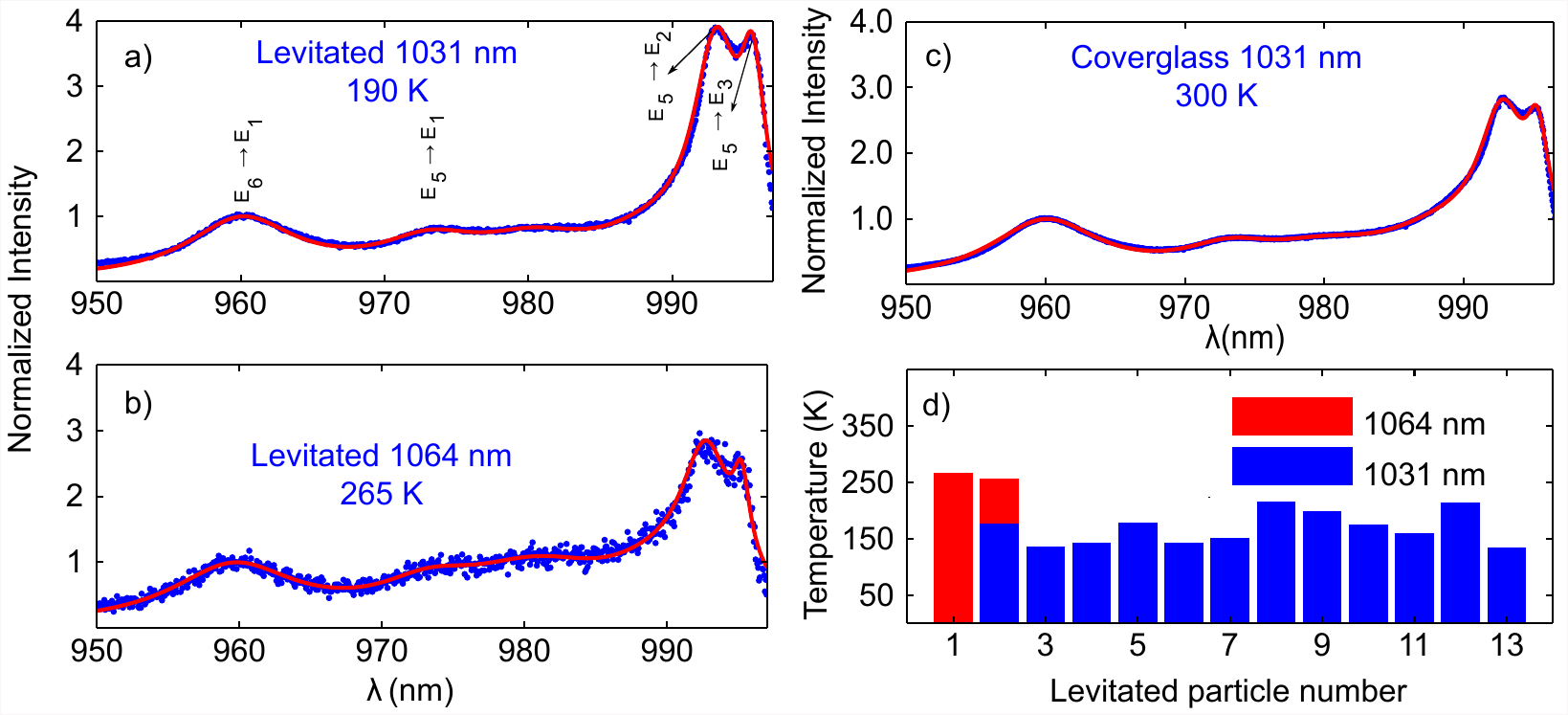}
\caption{ 
\footnotesize {\textbf{Blue shifted fluorescence from Yb$^{3+}$:YLF nanocrystals in $1031$ nm and $1064$ nm trapping beams.}} Red solid lines are fits considering six dominant transitions of Yb ions in the YLF host. These transitions are modeled as contributions from $E_6 \rightarrow E_1$, $E_5 \rightarrow E_1$, $E_6 \rightarrow E_2$, $E_6\rightarrow E_3$, $E_5 \rightarrow E_2$, and $E_5 \rightarrow E_3$ transitions, respectively. Spectra from levitated nanocrystals 
\textbf{ a)} using $1031$ nm and \textbf{ b)} $1064$ nm wavelength levitation. 
\textbf{ c)} Fluorescence spectrum from a Yb$^{3+}$:YLF nanocrystal when deposited on a $SiO_2$ coverglass. Temperatures are determined using the excited states population distribution assuming that particles are at room temperature when deposited on a coverglass. \textbf{ d)} A collection of levitated particles and their internal temperatures upon $1031$ nm and $1064$ nm levitation. Particle $2$ was initially trapped using $1031$ nm beam and then was gradually transferred to $1064$ nm beam while keeping everything else same. In $1031$ nm trapping $T$ was $\approx 175$ K while that in $1064$ nm was $255$ K. Fluorescence spectrum of particle number $5$ is shown in panel \textbf{a}. In this case $T$ is equal to $190$ K.}  
\label{fig2}  
\end{figure*}

The Yb$^{3+}$:YLF crystal temperature is determined from the relative population of the crystal-field split states E$_5$, E$_6$ and E$_7$ in the upper $F_{5/2}$ electronic state via the Boltzmann distribution assuming thermal equilibrium. Fluorescence spectra were collected through the objective lens also used for trapping. We fit spectral peaks to the recorded fluorescence profile which takes into account the major transitions within this observed region. Identification of each transition, and the energy of the levels in each electronic state, are taken from the literature \cite{Bensalah2004}. We assume Lorentzian profiles for each transition and that there are no strong vibronic transitions within the $950$ nm to $1000$ nm spectral region that we use to determine the temperature. Figures \ref{fig2}a and b shows typical spectra from levitated Yb$^{3+}$:YLF nanocrystals at laser wavelengths of $1031$ nm and $1064$ nm, respectively. Each spectrum is normalised to the $E_6 \rightarrow E_1$ transition peak at $960$ nm in order to compare the relative intensity of the $E_5 \rightarrow E_1$, $E_5 \rightarrow E_2$, and $E_5 \rightarrow E_3$ transitions. It can be clearly seen that the relative heights of transitions originating from E$_5$ level are higher for the $1031$ nm levitation when compared with the $1064$ nm spectrum. This indicates that the $1031$ nm levitated crystal is at a significantly lower temperature than the crystal levitated with $1064$ nm light. Figure \ref{fig2}c is the spectrum obtained from a nanocrystal that is deposited on a SiO$_2$ coverglass. It is also illuminated by the $1031$ nm laser with the same power used for levitation. Although many crystals are deposited on the coverglass we record the spectrum of one with its c-axis parallel to the laser polarization as determined by its spectral profile. We assume that the crystal is at room temperature as it is directly in contact with the cover slip. This was confirmed by reducing the laser power by a factor of ten which shows no change in the relative heights of the spectral peaks.

To quantify the temperatures we use the spectral fits (solid red lines in Fig. \ref{fig2}a$\&$b), at each trapping wavelength, to isolate the three peaks of interest corresponding to transitions $E_5 \rightarrow E_1$ ($972$ nm), $E_5 \rightarrow E_3$ ($993$ nm), and $E_6 \rightarrow E_1$ ($960$ nm). We use the Boltzmann relation to determine the change in temperature with respect to the crystal on the cover slip.  Assuming that the only change in spectral intensity is due to population differences in the upper levels, E$_5$, E$_6$ and E$_7$, and that these states are in thermal equilibrium, the change in the ratio of the peak areas for each transition can be used to determine the relative population of the E$_5$ and E$_6$ levels and hence to find the temperature. Importantly, as we are comparing the same transition at different temperatures, the transition probabilities do not need to be known to determine the temperature from these relative heights provided a reference spectrum at temperature, $T_0$ is available. For the reference we use a spectrum recorded from the crystal deposited on the cover slip and illuminated with the 1031 nm laser ($T_0=295$ K). This method has been used with rare earth crystals as a calibrated temperature probe\cite{Maurice1997,Collins1998,Wade2003,HaroGonzález2011, McLaurin2013,Wang2015,Li2016}. The temperature of the crystal $T$ can be determined from 

\begin{eqnarray}
\frac{1}{T}=\frac{1}{T_0}+\frac{k_B}{\Delta E_{65}}\ln{\frac{R_0}{R}},
\label{eqn0}
\end{eqnarray}

where $R$ is the intensity area ratio of well defined transitions $E_6 \rightarrow E_1$ and $E_5 \rightarrow E_1$ or $E_6 \rightarrow E_1$ and $E_5 \rightarrow E_2$ determined from the fits at temperature $T$, while $R_0$ is the same ratio at the room reference temperature taken to be $T_0=295$ K from the cover slip data. Additionally, $\Delta E_{65}=E_6-E_5$, is the energy difference between the E$_6$ and E$_5$ states, and $k_B$ is the Boltzmann's constant.

Figure \ref{fig2}d shows the temperature determined by this method for $13$ particles using $1031$ nm and $1064$ nm trapping light. For each spectrum we average the two temperatures obtained from the intensity area ratios $\frac{A(E_6 \rightarrow E_1)}{A(E_5 \rightarrow E_1)}$ and $\frac{A(E_6 \rightarrow E_2)}{A (E_5  \rightarrow E_1)}$. Although there is some variation between the temperature of particles trapped by 1031 nm light we find an average internal temperature of $167$ K. For the fewer particles trapped at $1064$ nm we find an average temperature of 260 K. Both of these numbers are consistent with the temperatures observed in bulk laser refrigeration in vacuum at these wavelengths and indicate that significantly lower temperatures will be obtained with trapping light near $1020$ nm where the minimum temperature has been achieved in bulk crystals. To compare the temperatures achievable on the same particle with different trapping wavelengths we initially trapped a particle using $1031$ nm light and recorded spectra. Subsequently, we moved the particle to the $1064$ nm trap by slowly decreasing the 1031 nm intensity while increasing the 1064 nm power. The temperature of this particle at both wavelengths is shown as particle number 2 in Fig. \ref{fig2}d.

\begin{figure*}
\includegraphics{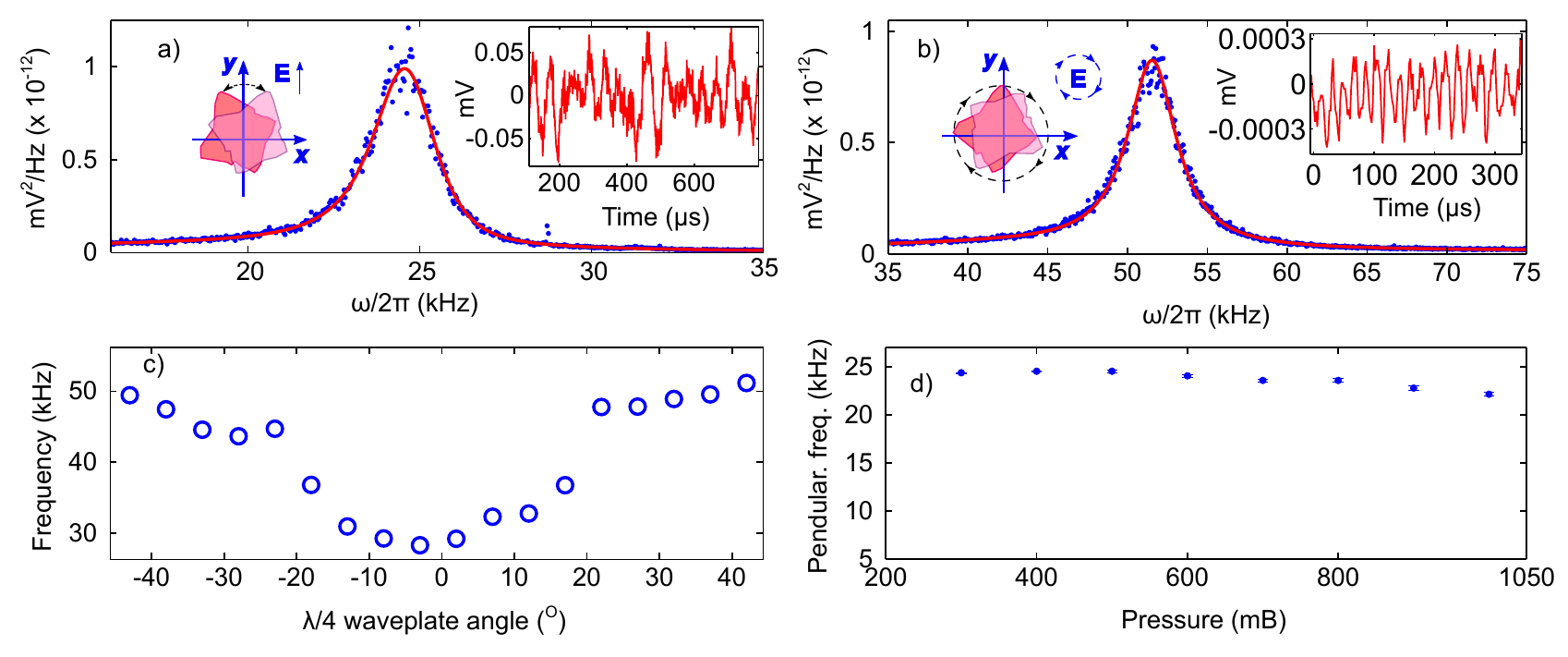}
\caption{
\footnotesize {\textbf{Alignment and rotation of levitated Yb$^{3+}$:YLF particles.}}
\textbf{ a)} Power spectral density (PSD) of a levitated Yb$^{3+}$:YLF particle at $\approx 500$ mB as it performs pendular motion in linearly polarized light. That's when an asymmetric particle is levitated in linearly polarized light, its long axis is aligned with the direction of polarization. Any misalignment due to collisions with gas molecules is counteracted by the laser field and particles start to perform harmonic oscillation. \textbf{ b)} PSD of a spinning levitated Yb$^{3+}$:YLF particle in circularly polarized (CP) light at $\approx 500$ mB. Due to the inherent asymmetry in the refractive indices along different axes of Yb$^{3+}$:YLF crystal, CP light will also impart torque and rotate the trapped particle. \textbf{ c)} Rotational frequency as a function of angle $\theta$ on the quarter waveplate. $\theta=0$ is equivalent to LP light while $|\theta|=45^o$ signifies CP light. $0<|\theta|<45$ produces elliptically polarized light and particles rotate at rates in between those at $\theta=0$ and $|\theta|=45^o$. \textbf{ d)} Pendular oscillation frequency as a function of pressure.}   
\label{fig3}  
\end{figure*}

The natural birefringence and shape induced birefringence of the asymmetric trapped Yb$^{3+}$:YLF nanocrystals allow us to observe rotation and alignment of the particles via the transfer of angular momentum from the light. Figure \ref{fig3}a shows a plot of the power spectral density (PSD) under the linearly polarised trapping light. The inset graph is a small part of the recorded oscillatory signal intensity that was used to produce the PSD at $\approx 500$ mB. The linearly polarized trapping light leads to a type of rotational simple harmonic motion (known as pendular or torsional motion) when the nanocrystal aligns itself with the polarisation axis of the trapping beam\cite{HoangPRL2016}. Interference between the trapping light and the Doppler shifted light induced by the particle leads to the observed modulation of the intensity when observed on a single diode. This motion is also observed on a split diode detector that is sensitive to the intensity modulation created by the rotation of an irregularly shaped particle at the trap focus. Although that the oscillation of the centre-of-mass motion of a particle in a trap also leads to similar signals in a split diode detector, at the higher pressures we use here, this motion is strongly damped and cannot be easily observed in the PSD. Figure \ref{fig3}b is a plot of the PSD recorded at the same pressure when the circularly polarized (CP) light is used to trap the particle. The inset shows a short time series of the intensity modulation due to the rotational Doppler shift. In this case, however, transfer of spin angular momentum to the particle, as originally observed in Beth's experiment \cite{Beth1936, Dunlop}, and more recently applied to levitated birefringent particles in vacuum\cite{Arita2013,Kuhn2016}, leads to a driven rotation of the trapped particle. We can smoothly tune between these two types of motion by changing the polarization of the trapping beam using a $\frac{\lambda}{4}$ waveplate. In addition to the usual rotational motion due to the inherent birefringence of Yb$^{3+}$:YLF crystal, absorption of CP light by Yb$^{3+}$ ions also transfers angular momentum to the weakly absorbing nanoparticle. Figure \ref{fig3}c is a plot of the observed frequency modulation of the transmitted trapping signal as a function of waveplate angle. When the waveplate is at $\pm$45 degrees the trapping beam is circularly polarised and linearly polarized at 0 degrees. At all other angles the beam is elliptically polarised. For rotational motion induced by circularly polarized light the rotation frequency increases for decreased gas damping. However, for linearly polarised light the frequency of the pendular motion does not change with damping as confirmed by the plot in Fig. \ref{fig3}d which shows that measured pendular frequency as a function of gas pressure is constant.

\begin{figure}
\includegraphics[width=0.5\textwidth]{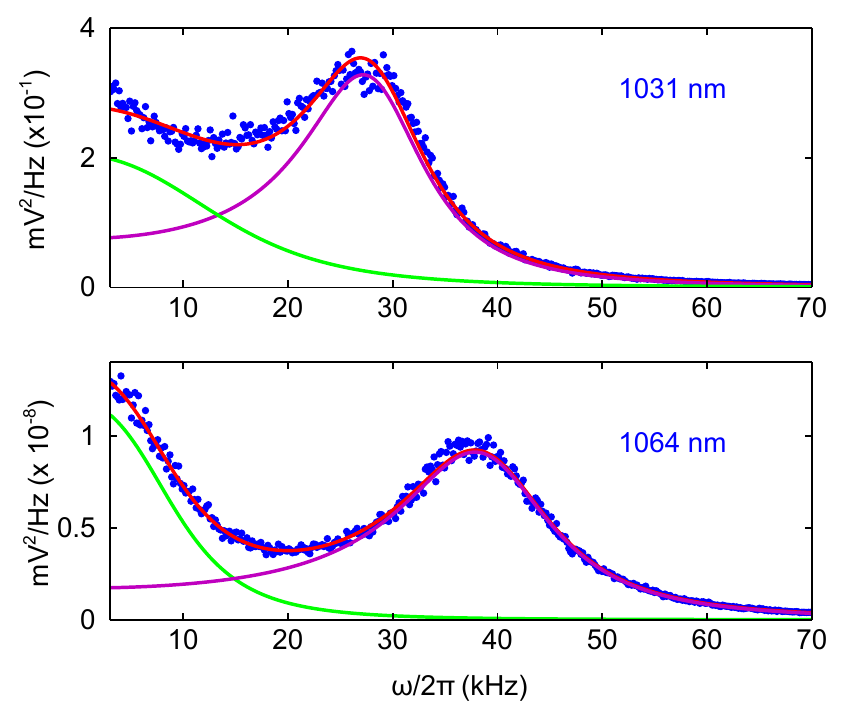}
\caption{\label{fig4} 
\footnotesize{\textbf{Power spectral density.}} PSDs corresponding to the pendular motion, oscillation of the elongated particles about the polarization axis of the laser's electric field, of particle number $2$ in Fig. \ref{fig2}d at $600$ mB \textbf{a)} in $1031$ nm, and \textbf{b)} in $1064$ nm trapping, respectively. The blue dots are data points while the red solid lines are fits corresponding to two harmonic oscillators - one for the translational motion (low frequency regime) and the other for the pendular oscillation (around the main peak). From the fit we retrieve damping $\Gamma$ that the particle encounters while oscillating in the $1031$ nm and $1064$ nm traps. As a complementary technique to the fluorescence intensity ratio based thermometry, we find $T$ from $\Gamma \infty \sqrt{T}$ (for more details see text).}    
\end{figure}

When the crystal is levitated in gas, refrigeration leads to cooling of the surrounding gas molecules and also to a modification in the local gas viscosity. The modified viscosity can be observed as a change in the viscous damping rate encountered by the levitated particles. For air at the temperatures and pressures we use, the viscosity and the resultant damping rate scales \cite{Blundell2006} with temperature as $T^{1/2}$. The change in the damping rate with temperature in different wavelength traps for the same particle can therefore be used to measure particle temperature. Figure \ref{fig4}a$\&$b are the PSDs of a single particle (number 2 in Fig. \ref{fig2}d) trapped in $1031$ nm and $1064$ nm wavelength light at a pressure of $600$ mB. The PSDs shown in the Fig. \ref{fig4} contain fits to the data that include both the translational degree of freedom (green, low freq. regime) and the pendular motion (purple, main peak). The damping rate derived from the fit to the pendular motion using 1031 nm light was found to be $\Gamma^p_{1031}/2\pi=13.94\pm 0.85$ kHz and $\Gamma^p_{1064}/2\pi=17.03\pm 0.48$ kHz for the 1064 nm trap giving a damping ratio of $R_p=\Gamma^p_{1031}/{\Gamma^p_{1064}}=0.82 \pm 0.05$.  For the translational motion we determine a damping ratio of  $R_t=0.74\pm0.1$. Assuming that the internal temperature of the particle in the $1064$ nm beam is $T_0=255$ K, obtained from our fluorescence measurements, we determine the temperature in the 1031 nm trap to be $T=T_0 R_p^2=171\pm 20$ K. From the damping of the translation motion we obtain a temperature of $140\pm 30$ K. Considering that a temperature gradient will exist around the particle and that the particles does not have a regular shape this simple model agrees well with the fluorescence ratio based prediction of $175$ K for the temperature in the 1031 nm trap. 

Finally, it is instructive to show that from simple considerations that the cooling power is sufficient to significantly reduce the temperature of the levitated crystals even in the presence of high gas pressures up to 500 mbar. If we consider the heating rate of a spherical particle due to the temperature difference, $\Delta T$, between the particle and the gas, the heat transfer rate\cite{Liu2006} is given approximately by $\dot{q_h}=4 \pi a k_g \Delta T,$ where $k_g$ is the heat conductivity of dry air, and $a$ is the particle radius. The cooling rate by laser refrigeration is given by $\dot{q_c}= \alpha I(1-\eta_q\frac{h \nu_f}{h \nu})$ where $\alpha$ is the absorption coefficient, $I$ is the laser intensity, and $\nu$ and $\nu_f$ are the excitation and average emission frequencies\cite{Epstein2009}. By equating these heating and cooling rates for a particle of radius 500 nm, we derive an equilibrium particle temperature of approximately 200 K using 1031 nm light. Despite the simplicity of the model it is consistent with the temperatures determined experimentally.

We have demonstrated laser refrigeration combined with the isolation provided by levitation to create a levitated nanocryostat whose orientation and rotation can be controlled via the particles' inherent birefringence. Lower temperatures should be achievable using a trapping laser at 1020 nm and the growth of high purity single nanocrystals\cite{Roder2015}.  Such a system is attractive for optical temperature control of single crystals without the need for traditional large cryostats. Yb$^{3+}$:YLF crystals can also in principle be grown around\cite{Neukirch2015} or bonded to other nanoparticles of interest such as nitrogen vacancy centres contained in diamond. This opens up the possibility for low temperature, high resolution nanoparticle spectroscopy and characterization. As cryogenic cooling is also key for increasing the spin coherence times in solid state systems it will extend the scope of future levitated hybrid optomechanics experiments where both translational and internal temperatures can now be controlled.



\begin{thebibliography}{34}
\expandafter\ifx\csname natexlab\endcsname\relax\def\natexlab#1{#1}\fi
\expandafter\ifx\csname bibnamefont\endcsname\relax
  \def\bibnamefont#1{#1}\fi
\expandafter\ifx\csname bibfnamefont\endcsname\relax
  \def\bibfnamefont#1{#1}\fi
\expandafter\ifx\csname citenamefont\endcsname\relax
  \def\citenamefont#1{#1}\fi
\expandafter\ifx\csname url\endcsname\relax
  \def\url#1{\texttt{#1}}\fi
\expandafter\ifx\csname urlprefix\endcsname\relax\def\urlprefix{URL }\fi
\providecommand{\bibinfo}[2]{#2}
\providecommand{\eprint}[2][]{\url{#2}}

\bibitem[{\citenamefont{Wan et~al.}(2016)\citenamefont{Wan, Scala, Morley,
  Rahman, Ulbricht, Bateman, Barker, Bose, and Kim}}]{WanPRL2016}
\bibinfo{author}{\bibfnamefont{C.}~\bibnamefont{Wan}},
  \bibinfo{author}{\bibfnamefont{M.}~\bibnamefont{Scala}},
  \bibinfo{author}{\bibfnamefont{G.~W.} \bibnamefont{Morley}},
  \bibinfo{author}{\bibfnamefont{A. T. M. ~A.} \bibnamefont{Rahman}},
  \bibinfo{author}{\bibfnamefont{H.}~\bibnamefont{Ulbricht}},
  \bibinfo{author}{\bibfnamefont{J.}~\bibnamefont{Bateman}},
  \bibinfo{author}{\bibfnamefont{P.~F.} \bibnamefont{Barker}},
  \bibinfo{author}{\bibfnamefont{S.}~\bibnamefont{Bose}}, \bibnamefont{and}
  \bibinfo{author}{\bibfnamefont{M.~S.} \bibnamefont{Kim}},
  \bibinfo{journal}{Phys. Rev. Lett.} \textbf{\bibinfo{volume}{117}},
  \bibinfo{pages}{143003} (\bibinfo{year}{2016}),
  

\bibitem[{\citenamefont{Scala et~al.}(2013)\citenamefont{Scala, Kim, Morley,
  Barker, and Bose}}]{Scala2013}
\bibinfo{author}{\bibfnamefont{M.}~\bibnamefont{Scala}},
  \bibinfo{author}{\bibfnamefont{M.~S.} \bibnamefont{Kim}},
  \bibinfo{author}{\bibfnamefont{G.~W.} \bibnamefont{Morley}},
  \bibinfo{author}{\bibfnamefont{P.~F.} \bibnamefont{Barker}},
  \bibnamefont{and} \bibinfo{author}{\bibfnamefont{S.}~\bibnamefont{Bose}},
  \bibinfo{journal}{Phys. Rev. Lett.} \textbf{\bibinfo{volume}{111}},
  \bibinfo{pages}{180403} (\bibinfo{year}{2013}),
  .

\bibitem[{\citenamefont{Yin et~al.}(2013)\citenamefont{Yin, Li, Zhang, and
  Duan}}]{Zhang2013}
\bibinfo{author}{\bibfnamefont{Z.-q.} \bibnamefont{Yin}},
  \bibinfo{author}{\bibfnamefont{T.}~\bibnamefont{Li}},
  \bibinfo{author}{\bibfnamefont{X.}~\bibnamefont{Zhang}}, \bibnamefont{and}
  \bibinfo{author}{\bibfnamefont{L.}~\bibnamefont{Duan}},
  \bibinfo{journal}{Phys. Rev. A} \textbf{\bibinfo{volume}{88}},
  \bibinfo{pages}{033614} (\bibinfo{year}{2013}),
  

\bibitem[{\citenamefont{Ranjit et~al.}(2016)\citenamefont{Ranjit, Cunningham,
  Casey, and Geraci}}]{GambhirPRA2016}
\bibinfo{author}{\bibfnamefont{G.}~\bibnamefont{Ranjit}},
  \bibinfo{author}{\bibfnamefont{M.}~\bibnamefont{Cunningham}},
  \bibinfo{author}{\bibfnamefont{K.}~\bibnamefont{Casey}}, \bibnamefont{and}
  \bibinfo{author}{\bibfnamefont{A.~A.} \bibnamefont{Geraci}},
  \bibinfo{journal}{Phys. Rev. A} \textbf{\bibinfo{volume}{93}},
  \bibinfo{pages}{053801} (\bibinfo{year}{2016}),

\bibitem[{\citenamefont{Gieseler et~al.}(2014)\citenamefont{Gieseler, Quidant,
  Dellago, and Novotny}}]{GieselerNatNano2014}
\bibinfo{author}{\bibfnamefont{J.}~\bibnamefont{Gieseler}},
  \bibinfo{author}{\bibfnamefont{R.}~\bibnamefont{Quidant}},
  \bibinfo{author}{\bibfnamefont{C.}~\bibnamefont{Dellago}}, \bibnamefont{and}
  \bibinfo{author}{\bibfnamefont{L.}~\bibnamefont{Novotny}},
  \bibinfo{journal}{Nat. Nano} \textbf{\bibinfo{volume}{9}},
  \bibinfo{pages}{358 } (\bibinfo{year}{2014}).

\bibitem[{\citenamefont{Millen et~al.}(2014)\citenamefont{Millen, Deesuwan,
  Barker, and Anders}}]{MillenNat2014}
\bibinfo{author}{\bibfnamefont{J.}~\bibnamefont{Millen}},
  \bibinfo{author}{\bibfnamefont{T.}~\bibnamefont{Deesuwan}},
  \bibinfo{author}{\bibfnamefont{P.}~\bibnamefont{Barker}}, \bibnamefont{and}
  \bibinfo{author}{\bibfnamefont{J.}~\bibnamefont{Anders}},
  \bibinfo{journal}{Nat. Nano.} \textbf{\bibinfo{volume}{9}},
  \bibinfo{pages}{425} (\bibinfo{year}{2014}).

\bibitem[{\citenamefont{Jain et~al.}(2016)\citenamefont{Jain, Gieseler, Moritz,
  Dellago, Quidant, and Novotny}}]{Jain2016}
\bibinfo{author}{\bibfnamefont{V.}~\bibnamefont{Jain}},
  \bibinfo{author}{\bibfnamefont{J.}~\bibnamefont{Gieseler}},
  \bibinfo{author}{\bibfnamefont{C.}~\bibnamefont{Moritz}},
  \bibinfo{author}{\bibfnamefont{C.}~\bibnamefont{Dellago}},
  \bibinfo{author}{\bibfnamefont{R.}~\bibnamefont{Quidant}}, \bibnamefont{and}
  \bibinfo{author}{\bibfnamefont{L.}~\bibnamefont{Novotny}},
  \bibinfo{journal}{Phys. Rev. Lett.} \textbf{\bibinfo{volume}{116}},
  \bibinfo{pages}{243601} (\bibinfo{year}{2016}).

\bibitem[{\citenamefont{Li et~al.}(2011)\citenamefont{Li, Kheifets, and
  Raizen}}]{LiNatPhys2011}
\bibinfo{author}{\bibfnamefont{T.}~\bibnamefont{Li}},
  \bibinfo{author}{\bibfnamefont{S.}~\bibnamefont{Kheifets}}, \bibnamefont{and}
  \bibinfo{author}{\bibfnamefont{M.~G.} \bibnamefont{Raizen}},
  \bibinfo{journal}{Nat. Phys.} \textbf{\bibinfo{volume}{7}},
  \bibinfo{pages}{527} (\bibinfo{year}{2011}).

\bibitem[{\citenamefont{Gieseler et~al.}(2012)\citenamefont{Gieseler, Deutsch,
  Quidant, and Novotny}}]{Gieseler2012}
\bibinfo{author}{\bibfnamefont{J.}~\bibnamefont{Gieseler}},
  \bibinfo{author}{\bibfnamefont{B.}~\bibnamefont{Deutsch}},
  \bibinfo{author}{\bibfnamefont{R.}~\bibnamefont{Quidant}}, \bibnamefont{and}
  \bibinfo{author}{\bibfnamefont{L.}~\bibnamefont{Novotny}},
  \bibinfo{journal}{Phys. Rev. Lett.} \textbf{\bibinfo{volume}{109}},
  \bibinfo{pages}{103603} (\bibinfo{year}{2012}).

\bibitem[{\citenamefont{Pringsheim}(1929)}]{Pringsheim1929}
\bibinfo{author}{\bibfnamefont{P.}~\bibnamefont{Pringsheim}},
  \bibinfo{journal}{Zeitschrift f{\"u}r Physik} \textbf{\bibinfo{volume}{57}},
  \bibinfo{pages}{739} (\bibinfo{year}{1929}), ISSN \bibinfo{issn}{0044-3328}.

\bibitem[{\citenamefont{Metcalf and van~der Straten}(2007)}]{Metcalf2007}
\bibinfo{author}{\bibfnamefont{H.~J.} \bibnamefont{Metcalf}} \bibnamefont{and}
  \bibinfo{author}{\bibfnamefont{P.}~\bibnamefont{van~der Straten}},
  \emph{\bibinfo{title}{Laser Cooling and Trapping of Neutral Atoms}}
  (\bibinfo{year}{2007}).

\bibitem[{\citenamefont{Mungan et~al.}(1997)\citenamefont{Mungan, Buchwald,
  Edwards, Epstein, and Gosnell}}]{Mungan1997}
\bibinfo{author}{\bibfnamefont{C.~E.} \bibnamefont{Mungan}},
  \bibinfo{author}{\bibfnamefont{M.~I.} \bibnamefont{Buchwald}},
  \bibinfo{author}{\bibfnamefont{B.~C.} \bibnamefont{Edwards}},
  \bibinfo{author}{\bibfnamefont{R.~I.} \bibnamefont{Epstein}},
  \bibnamefont{and} \bibinfo{author}{\bibfnamefont{T.~R.}
  \bibnamefont{Gosnell}}, \bibinfo{journal}{Phys. Rev. Lett.}
  \textbf{\bibinfo{volume}{78}}, \bibinfo{pages}{1030} (\bibinfo{year}{1997}).

\bibitem[{\citenamefont{Seletskiy et~al.}(2010)\citenamefont{Seletskiy,
  Melgaard, Bigotta, Lieto, Tonelli, and Sheik-Bahae}}]{Seletskiy2010}
\bibinfo{author}{\bibfnamefont{D.~V.} \bibnamefont{Seletskiy}},
  \bibinfo{author}{\bibfnamefont{S.~D.} \bibnamefont{Melgaard}},
  \bibinfo{author}{\bibfnamefont{S.}~\bibnamefont{Bigotta}},
  \bibinfo{author}{\bibfnamefont{A.~D.} \bibnamefont{Lieto}},
  \bibinfo{author}{\bibfnamefont{M.}~\bibnamefont{Tonelli}}, \bibnamefont{and}
  \bibinfo{author}{\bibfnamefont{M.}~\bibnamefont{Sheik-Bahae}},
  \bibinfo{journal}{Nat. Phot.} \textbf{\bibinfo{volume}{4}},
  \bibinfo{pages}{161} (\bibinfo{year}{2010}).

\bibitem[{\citenamefont{Melgaard et~al.}(2016)\citenamefont{Melgaard, Albrecht,
  Hehlen, and Sheik-Bahae}}]{Melgaard2016}
\bibinfo{author}{\bibfnamefont{S.~D.} \bibnamefont{Melgaard}},
  \bibinfo{author}{\bibfnamefont{A.~R.} \bibnamefont{Albrecht}},
  \bibinfo{author}{\bibfnamefont{M.~P.} \bibnamefont{Hehlen}},
  \bibnamefont{and}
  \bibinfo{author}{\bibfnamefont{M.}~\bibnamefont{Sheik-Bahae}},
  \bibinfo{journal}{Sci. Rep.} \textbf{\bibinfo{volume}{6}},
  \bibinfo{pages}{20380} (\bibinfo{year}{2016}).

\bibitem[{\citenamefont{Epstein et~al.}(1995)\citenamefont{Epstein, Buchwald,
  Edwards, Gosnell, and Mungan}}]{Epstein1995}
\bibinfo{author}{\bibfnamefont{R.~I.} \bibnamefont{Epstein}},
  \bibinfo{author}{\bibfnamefont{M.~I.} \bibnamefont{Buchwald}},
  \bibinfo{author}{\bibfnamefont{B.~C.} \bibnamefont{Edwards}},
  \bibinfo{author}{\bibfnamefont{T.~R.} \bibnamefont{Gosnell}},
  \bibnamefont{and} \bibinfo{author}{\bibfnamefont{C.~E.}
  \bibnamefont{Mungan}}, \bibinfo{journal}{Nature}
  \textbf{\bibinfo{volume}{377}}, \bibinfo{pages}{500} (\bibinfo{year}{1995}).

\bibitem[{\citenamefont{Roder et~al.}(2015)\citenamefont{Roder, Smith, Zhou,
  Crane, and Pauzauskie}}]{Roder2015}
\bibinfo{author}{\bibfnamefont{P.~B.} \bibnamefont{Roder}},
  \bibinfo{author}{\bibfnamefont{B.~E.} \bibnamefont{Smith}},
  \bibinfo{author}{\bibfnamefont{X.}~\bibnamefont{Zhou}},
  \bibinfo{author}{\bibfnamefont{M.~J.} \bibnamefont{Crane}}, \bibnamefont{and}
  \bibinfo{author}{\bibfnamefont{P.~J.} \bibnamefont{Pauzauskie}},
  \bibinfo{journal}{PNAS} \textbf{\bibinfo{volume}{112}},
  \bibinfo{pages}{15024} (\bibinfo{year}{2015}).

\bibitem[{\citenamefont{Bensalah et~al.}(2004)\citenamefont{Bensalah, Guyot,
  Ito, Brenier, Sato, Fukuda, and Boulon}}]{Bensalah2004}
\bibinfo{author}{\bibfnamefont{A.}~\bibnamefont{Bensalah}},
  \bibinfo{author}{\bibfnamefont{Y.}~\bibnamefont{Guyot}},
  \bibinfo{author}{\bibfnamefont{M.}~\bibnamefont{Ito}},
  \bibinfo{author}{\bibfnamefont{A.}~\bibnamefont{Brenier}},
  \bibinfo{author}{\bibfnamefont{H.}~\bibnamefont{Sato}},
  \bibinfo{author}{\bibfnamefont{T.}~\bibnamefont{Fukuda}}, \bibnamefont{and}
  \bibinfo{author}{\bibfnamefont{G.}~\bibnamefont{Boulon}},
  \bibinfo{journal}{Opt. Mater.} \textbf{\bibinfo{volume}{26}},
  \bibinfo{pages}{375 } (\bibinfo{year}{2004}).

\bibitem[{\citenamefont{Sugiyama et~al.}(2006)\citenamefont{Sugiyama,
  Katsurayama, Anzai, and Tsuboi}}]{Sugiyama2006}
\bibinfo{author}{\bibfnamefont{A.}~\bibnamefont{Sugiyama}},
  \bibinfo{author}{\bibfnamefont{M.}~\bibnamefont{Katsurayama}},
  \bibinfo{author}{\bibfnamefont{Y.}~\bibnamefont{Anzai}}, \bibnamefont{and}
  \bibinfo{author}{\bibfnamefont{T.}~\bibnamefont{Tsuboi}},
  \bibinfo{journal}{J. Alloys Compd.}
  \textbf{\bibinfo{volume}{408–412}}, \bibinfo{pages}{780 }
  (\bibinfo{year}{2006}).

\bibitem[{\citenamefont{Fan et~al.}(2007)\citenamefont{Fan, Ripin, Aggarwal,
  Ochoa, Chann, Tilleman, and Spitzberg}}]{Fan2007}
\bibinfo{author}{\bibfnamefont{T.~Y.} \bibnamefont{Fan}},
  \bibinfo{author}{\bibfnamefont{D.~J.} \bibnamefont{Ripin}},
  \bibinfo{author}{\bibfnamefont{R.~L.} \bibnamefont{Aggarwal}},
  \bibinfo{author}{\bibfnamefont{J.~R.} \bibnamefont{Ochoa}},
  \bibinfo{author}{\bibfnamefont{B.}~\bibnamefont{Chann}},
  \bibinfo{author}{\bibfnamefont{M.}~\bibnamefont{Tilleman}}, \bibnamefont{and}
  \bibinfo{author}{\bibfnamefont{J.}~\bibnamefont{Spitzberg}},
  \bibinfo{journal}{IEEE J. Sel. Top. Quantum Electron.}
  \textbf{\bibinfo{volume}{13}}, \bibinfo{pages}{448} (\bibinfo{year}{2007}).

\bibitem[{\citenamefont{Seletskiy et~al.}(2013)\citenamefont{Seletskiy,
  Melgaard, Epstein, Lieto, Tonelli, and Sheik-Bahae}}]{Seletskiy2013}
\bibinfo{author}{\bibfnamefont{D.~V.} \bibnamefont{Seletskiy}},
  \bibinfo{author}{\bibfnamefont{S.~D.} \bibnamefont{Melgaard}},
  \bibinfo{author}{\bibfnamefont{R.~I.} \bibnamefont{Epstein}},
  \bibinfo{author}{\bibfnamefont{A.~D.} \bibnamefont{Lieto}},
  \bibinfo{author}{\bibfnamefont{M.}~\bibnamefont{Tonelli}}, \bibnamefont{and}
  \bibinfo{author}{\bibfnamefont{M.}~\bibnamefont{Sheik-Bahae}},
  \bibinfo{journal}{J. Lumin.} \textbf{\bibinfo{volume}{133}},
  \bibinfo{pages}{5 } (\bibinfo{year}{2013}).

\bibitem[{\citenamefont{Singer et~al.}(2006)\citenamefont{Singer, Nieminen,
  Gibson, Heckenberg, and Rubinsztein-Dunlop}}]{Singer2006}
\bibinfo{author}{\bibfnamefont{W.}~\bibnamefont{Singer}},
  \bibinfo{author}{\bibfnamefont{T.~A.} \bibnamefont{Nieminen}},
  \bibinfo{author}{\bibfnamefont{U.~J.} \bibnamefont{Gibson}},
  \bibinfo{author}{\bibfnamefont{N.~R.} \bibnamefont{Heckenberg}},
  \bibnamefont{and}
  \bibinfo{author}{\bibfnamefont{H.}~\bibnamefont{Rubinsztein-Dunlop}},
  \bibinfo{journal}{Phys. Rev. E} \textbf{\bibinfo{volume}{73}},
  \bibinfo{pages}{021911} (\bibinfo{year}{2006}).

\bibitem[{\citenamefont{Maurice et~al.}(1997)\citenamefont{Maurice, Wade,
  Collins, Monnom, and Baxter}}]{Maurice1997}
\bibinfo{author}{\bibfnamefont{E.}~\bibnamefont{Maurice}},
  \bibinfo{author}{\bibfnamefont{S.~A.} \bibnamefont{Wade}},
  \bibinfo{author}{\bibfnamefont{S.~F.} \bibnamefont{Collins}},
  \bibinfo{author}{\bibfnamefont{G.}~\bibnamefont{Monnom}}, \bibnamefont{and}
  \bibinfo{author}{\bibfnamefont{G.~W.} \bibnamefont{Baxter}},
  \bibinfo{journal}{Appl. Opt.} \textbf{\bibinfo{volume}{36}},
  \bibinfo{pages}{8264} (\bibinfo{year}{1997}).

\bibitem[{\citenamefont{Collins et~al.}(1998)\citenamefont{Collins, Baxter,
  Wade, Sun, Grattan, Zhang, and Palmer}}]{Collins1998}
\bibinfo{author}{\bibfnamefont{S.~F.} \bibnamefont{Collins}},
  \bibinfo{author}{\bibfnamefont{G.~W.} \bibnamefont{Baxter}},
  \bibinfo{author}{\bibfnamefont{S.~A.} \bibnamefont{Wade}},
  \bibinfo{author}{\bibfnamefont{T.}~\bibnamefont{Sun}},
  \bibinfo{author}{\bibfnamefont{K.~T.~V.} \bibnamefont{Grattan}},
  \bibinfo{author}{\bibfnamefont{Z.~Y.} \bibnamefont{Zhang}}, \bibnamefont{and}
  \bibinfo{author}{\bibfnamefont{A.~W.} \bibnamefont{Palmer}},
  \bibinfo{journal}{J. Appl. Phys.} \textbf{\bibinfo{volume}{84}},
  \bibinfo{pages}{4649} (\bibinfo{year}{1998}).

\bibitem[{\citenamefont{Wade et~al.}(2003)\citenamefont{Wade, Collins, and
  Baxter}}]{Wade2003}
\bibinfo{author}{\bibfnamefont{S.~A.} \bibnamefont{Wade}},
  \bibinfo{author}{\bibfnamefont{S.~F.} \bibnamefont{Collins}},
  \bibnamefont{and} \bibinfo{author}{\bibfnamefont{G.~W.}
  \bibnamefont{Baxter}}, \bibinfo{journal}{J. Appl. Phys.}
  \textbf{\bibinfo{volume}{94}}, \bibinfo{pages}{4743} (\bibinfo{year}{2003}).

\bibitem[{\citenamefont{Haro-González
  et~al.}(2011)\citenamefont{Haro-González, León-Luis, González-Pérez, and
  Martín}}]{HaroGonzález2011}
\bibinfo{author}{\bibfnamefont{P.}~\bibnamefont{Haro-González}},
  \bibinfo{author}{\bibfnamefont{S.}~\bibnamefont{León-Luis}},
  \bibinfo{author}{\bibfnamefont{S.}~\bibnamefont{González-Pérez}},
  \bibnamefont{and} \bibinfo{author}{\bibfnamefont{I.}~\bibnamefont{Martín}},
  \bibinfo{journal}{Mater. Res. Bull.} \textbf{\bibinfo{volume}{46}},
  \bibinfo{pages}{1051 } (\bibinfo{year}{2011}).

\bibitem[{\citenamefont{McLaurin et~al.}(2013)\citenamefont{McLaurin, Bradshaw,
  and Gamelin}}]{McLaurin2013}
\bibinfo{author}{\bibfnamefont{E.~J.} \bibnamefont{McLaurin}},
  \bibinfo{author}{\bibfnamefont{L.~R.} \bibnamefont{Bradshaw}},
  \bibnamefont{and} \bibinfo{author}{\bibfnamefont{D.~R.}
  \bibnamefont{Gamelin}}, \bibinfo{journal}{Chem. Mater.}
  \textbf{\bibinfo{volume}{25}}, \bibinfo{pages}{1283} (\bibinfo{year}{2013}).

\bibitem[{\citenamefont{Wang et~al.}(2015)\citenamefont{Wang, Liu, Bu, Liu,
  Liu, and Yan}}]{Wang2015}
\bibinfo{author}{\bibfnamefont{X.}~\bibnamefont{Wang}},
  \bibinfo{author}{\bibfnamefont{Q.}~\bibnamefont{Liu}},
  \bibinfo{author}{\bibfnamefont{Y.}~\bibnamefont{Bu}},
  \bibinfo{author}{\bibfnamefont{C.-S.} \bibnamefont{Liu}},
  \bibinfo{author}{\bibfnamefont{T.}~\bibnamefont{Liu}}, \bibnamefont{and}
  \bibinfo{author}{\bibfnamefont{X.}~\bibnamefont{Yan}}, \bibinfo{journal}{RSC
  Adv.} \textbf{\bibinfo{volume}{5}}, \bibinfo{pages}{86219}.

\bibitem[{\citenamefont{Li et~al.}(2016)\citenamefont{Li, Wang, Li, Zhang, and
  Yang}}]{Li2016}
\bibinfo{author}{\bibfnamefont{H.}~\bibnamefont{Li}},
  \bibinfo{author}{\bibfnamefont{Y.}~\bibnamefont{Wang}},
  \bibinfo{author}{\bibfnamefont{H.}~\bibnamefont{Li}},
  \bibinfo{author}{\bibfnamefont{Y.}~\bibnamefont{Zhang}}, \bibnamefont{and}
  \bibinfo{author}{\bibfnamefont{J.}~\bibnamefont{Yang}},
  \bibinfo{journal}{Sci. Rep.} \textbf{\bibinfo{volume}{6}}
  (\bibinfo{year}{2016}).

\bibitem[{\citenamefont{Hoang et~al.}(2016)\citenamefont{Hoang, Ma, Ahn, Bang,
  Robicheaux, Yin, and Li}}]{HoangPRL2016}
\bibinfo{author}{\bibfnamefont{T.~M.} \bibnamefont{Hoang}},
  \bibinfo{author}{\bibfnamefont{Y.}~\bibnamefont{Ma}},
  \bibinfo{author}{\bibfnamefont{J.}~\bibnamefont{Ahn}},
  \bibinfo{author}{\bibfnamefont{J.}~\bibnamefont{Bang}},
  \bibinfo{author}{\bibfnamefont{F.}~\bibnamefont{Robicheaux}},
  \bibinfo{author}{\bibfnamefont{Z.-Q.} \bibnamefont{Yin}}, \bibnamefont{and}
  \bibinfo{author}{\bibfnamefont{T.}~\bibnamefont{Li}}, \bibinfo{journal}{Phys.
  Rev. Lett.} \textbf{\bibinfo{volume}{117}}, \bibinfo{pages}{123604}
  (\bibinfo{year}{2016}).

\bibitem[{\citenamefont{Beth}(1936)}]{Beth1936}
\bibinfo{author}{\bibfnamefont{R.~A.} \bibnamefont{Beth}},
  \bibinfo{journal}{Phys. Rev.} \textbf{\bibinfo{volume}{50}},
  \bibinfo{pages}{115} (\bibinfo{year}{1936}).
  
 \bibitem[{\citenamefont{Friese et~al.}(1998)\citenamefont{Friese, Nieminen,
  Heckenberg, and Rubinsztein-Dunlop}}]{Dunlop}
\bibinfo{author}{\bibfnamefont{M.~E.~J.} \bibnamefont{Friese}},
  \bibinfo{author}{\bibfnamefont{T.~A.} \bibnamefont{Nieminen}},
  \bibinfo{author}{\bibfnamefont{N.~R.} \bibnamefont{Heckenberg}},
  \bibnamefont{and}
  \bibinfo{author}{\bibfnamefont{H.}~\bibnamefont{Rubinsztein-Dunlop}},
  \bibinfo{journal}{Nature} \textbf{\bibinfo{volume}{394}},
  \bibinfo{pages}{348} (\bibinfo{year}{1998}). 
  
\bibitem[{\citenamefont{Arita et~al.}(2013)\citenamefont{Arita, Mazilu, and
  Dholakia}}]{Arita2013}
\bibinfo{author}{\bibfnamefont{Y.}~\bibnamefont{Arita}},
  \bibinfo{author}{\bibfnamefont{M.}~\bibnamefont{Mazilu}}, \bibnamefont{and}
  \bibinfo{author}{\bibfnamefont{K.}~\bibnamefont{Dholakia}},
  \bibinfo{journal}{Nat. Comm.} \textbf{\bibinfo{volume}{4}},
  \bibinfo{pages}{2374} (\bibinfo{year}{2013}).
  
  
  \bibitem[{\citenamefont{Kuhn et~al.}(2016)\citenamefont{Kuhn, Kosloff,
  Stickler, Patolsky, Hornberger, Arndt, and Millen}}]{Kuhn2016}
\bibinfo{author}{\bibfnamefont{S.}~\bibnamefont{Kuhn}},
  \bibinfo{author}{\bibfnamefont{A.}~\bibnamefont{Kosloff}},
  \bibinfo{author}{\bibfnamefont{B.~A.} \bibnamefont{Stickler}},
  \bibinfo{author}{\bibfnamefont{F.}~\bibnamefont{Patolsky}},
  \bibinfo{author}{\bibfnamefont{K.}~\bibnamefont{Hornberger}},
  \bibinfo{author}{\bibfnamefont{M.}~\bibnamefont{Arndt}}, \bibnamefont{and}
  \bibinfo{author}{\bibfnamefont{J.}~\bibnamefont{Millen}},
  \bibinfo{journal}{arxiv}  (\bibinfo{year}{2016}).  

\bibitem[{\citenamefont{Blundell and Blundell}(2006)}]{Blundell2006}
\bibinfo{author}{\bibfnamefont{S.}~\bibnamefont{Blundell}} \bibnamefont{and}
  \bibinfo{author}{\bibfnamefont{C.}~\bibnamefont{Blundell}}
  (\bibinfo{year}{2006}).

\bibitem[{\citenamefont{Liu et~al.}(2006)\citenamefont{Liu, Daun, Snelling, and
  Smallwood}}]{Liu2006}
\bibinfo{author}{\bibfnamefont{F.}~\bibnamefont{Liu}},
  \bibinfo{author}{\bibfnamefont{K.}~\bibnamefont{Daun}},
  \bibinfo{author}{\bibfnamefont{D.}~\bibnamefont{Snelling}}, \bibnamefont{and}
  \bibinfo{author}{\bibfnamefont{G.}~\bibnamefont{Smallwood}},
  \bibinfo{journal}{Appl. Phys. B} \textbf{\bibinfo{volume}{83}},
  \bibinfo{pages}{355} (\bibinfo{year}{2006}).

\bibitem[{\citenamefont{Epstein and Sheik-Bahae}(2009)}]{Epstein2009}
\bibinfo{author}{\bibfnamefont{R.}~\bibnamefont{Epstein}} \bibnamefont{and}
  \bibinfo{author}{\bibfnamefont{M.}~\bibnamefont{Sheik-Bahae}},
  \emph{\bibinfo{title}{Optical Refrigeration in Solids: Fundamentals and
  Overview}} (\bibinfo{year}{2009}), pp. \bibinfo{pages}{1--32}.

\bibitem[{\citenamefont{Neukirch et~al.}(2015)\citenamefont{Neukirch, von
  Haartman, Rosenholm, and Vamivakas}}]{Neukirch2015}
\bibinfo{author}{\bibfnamefont{L.~P.} \bibnamefont{Neukirch}},
  \bibinfo{author}{\bibfnamefont{E.}~\bibnamefont{von Haartman}},
  \bibinfo{author}{\bibfnamefont{J.~M.} \bibnamefont{Rosenholm}},
  \bibnamefont{and}
  \bibinfo{author}{\bibfnamefont{N.}~\bibnamefont{Vamivakas}},
  \bibinfo{journal}{Nat. Photon.} \textbf{\bibinfo{volume}{9}},
  \bibinfo{pages}{653–657} (\bibinfo{year}{2015}).

\end{thebibliography}



\section*{Acknowledgements}  
\noindent We acknowledge support from the UK's Engineering and Physical Science Research Council grant. We also acknowledge the help of G. W. Morley and A. C. Frangeskou for Scanning Electron Microscopy imaging.

\section*{Author contributions} 
\noindent All authors contributed to the data analysis and wrote the manuscript.

\section*{Competing financial interests} 
\noindent The authors declare no competing financial interests. 

\newpage
\section*{Methods}
Nanocrystals are milled from a commercial laser $10\%$ Yb$^{3+}$:YLF crystal. To prepare the particles for levitation they were dispersed in HPLC grade methanol using an ultrasonic bath and introduced into the trapping volume by an ultrasonic nebuliser. Once a particle was trapped, the vacuum chamber was evacuated down to pressures as low as 20 mbar. Fluorescence from levitated particles, as well as from particles deposited on the cover-glass, were collected through the same objective lens used for trapping. Spectra were recorded using either a home built spectrometer consisting of an Andor Ltd EMCCD iXon $885$ camera detector and a $600$ lines per mm grating or using an Andor Ltd SR163 spectrometer equipped with a CCD detector. The pendular and rotational motion were detected on a set of balanced photodiodes (PDB210C from Thorlabs Ltd). A long time series of this data from the photodiodes was recorded on an oscilloscope and was post processed to extract the power spectral density. The pendular motion and the transverse trap frequency are simultaneously recorded. Fits to the data, assuming damped translational and pendular motion, were used to extract the damping ratios for both types of motion as a function of trapping wavelength. This data was used to confirm the temperatures were consistent with those obtained via the fluorescence measurements.       

\end{document}